\documentstyle[12pt,epsf]{article}

\begin{document}
\bibliographystyle{unsrt} 
\title{POLARIZATION AND STRUCTURE OF SMALL CLUSTERS.\\
}
\author{W. GEERTSMA, \\
Departamento de Fisica, UFES, Av. Fernando
Ferrari s/n\\
Vitoria--ES, Brasil.}
\maketitle

\begin{abstract}
In this paper   we report  the structure  of the Li$_{4}$Ge$_{4}$
cluster as a function  of charge transfer  and polarizability.  We
find that for small charge transfer ($Q < 0.5$)   this cluster has
the  expected cubic structure: a Ge$_4$ tetrahedron with  a Li ion
attached at  large distance to each face.   With increasing charge
transfer $Q>0.5$ the structure of Li$_{4}$Ge$_{4}$   changes:  for
relative small polarizability  of the Ge ion the Ge$_4$  breaks up
into two   Ge$_2$ pairs  separated  by  the  four  Li. For  larger
polarizability  there are three possibilities: 1)  for full charge
transfer  the Li  ions  break  up  the   Ge subcluster  into  four
separated ions; 2) for smaller   values of the charge transfer  we
still have  the structure  with  two  Ge$_2$  pairs    and  3) for
intermediate values  of, charge transfer the Ge sublattice forms a
structure  with two opposite  bonds  of the  Ge$_4$    tetrahedron
cluster broken.  These are the only stable  geometries found.  For
large Ge polarizability    we  find  that all   structures  become
unstable: the    size of the induced dipole  moment becomes larger
than  the  diameter   of Ge.  Based  on this   phase   diagram  of
Li$_{4}$Ge$_{4}$   we  discuss the  structure of other  A$_4$M$_4$
alkali(=  A)--tetralide  (= M) (= group 14)  clusters, and related
solid state  structures. 
\vfill
To be Submitted to J. Condens. Matter
\end{abstract}


\vfill

\newpage
\section{INTRODUCTION.}\label{section: introduction}
In this paper we study the influence   of   polarization   on  the
structure  of  small   ionic-covalent   clusters.  In  a  previous
paper~\cite{geertsma   2002} (to  be  referred  to  as I) on this
subject we have reported that polarizability   can be an important
factor  determining  the geometry of clusters, especially clusters
consisting   of electropositive  and  electronegative atoms.    We
illustrated   this by   the   calculation  of  the  structure   of
A$_4$M$_4$  clusters where  A is an alkali  ion: Li, Na, K, Rb, Cs
and M is a tetralide    ion:  Si, Ge, Sn,  Pb as a function of the
polarizability.     We  used  the full ionic   charges   in  these
calculations.    We find  that the structures  are only stable for
polarizabilities    smaller        than    about   0.7 times   the
Fraga~\cite{Fraga} values.    Secondly,  we  find that  for   this
maximum    value  of the polarizability the stable  structure   of
K$_4$M$_4$, Rb$_4$M$_4$  and  Cs$_4$M$_4$ is  a M$_4$  tetrahedron
with the  alkali   far outside the four faces of this tetrahedron:
the Normal Double  Tetrahedron (NDT) structure.   We found that in
case     of Li  and Na for small  polarizability    of  M the same
structure as for the other alkalis,  while for large values of the
polarizability  we found the Li or Na  near the faces of the M$_4$
tetrahedron:  the near Face Centered  Double     Tetrahedron (FDT)
structure.  For intermediate values of the polarizability  we find
two pairs    of M$_2$,  separated by  Li or Na (to be indicated by
PAR) and a boat or butterfly  like structure: a tetrahedron   with
two elongated opposite M--M bonds, we indicate  by BUT.  Note that
NDT  and   FDT  have $T_d$ symmetry, and PAR and  BUT   have $S_4$
symmetry.   One can  continuously  distort the first structures to
get the latter structures,  $T_d$ is a special   case    of  $S_4$
symmetry.

In the present paper we  will discuss these  clusters in some more
detail, with especial  emphasis of the dependence of the  geometry
on the ionic charge and  polarizability.  In  the calculations  we
reported in I we also found sometimes  a structure with the M ions
in a sort of   butterfly geometry (indicated by BUT).  However  we
always found this geometry to have  a smaller  binding energy than
the other geometries  for the M subcluster: two M$_2$ pairs (PAR),
the Normal Tetrahedron (NDT)  and  Near Face Centered  Tetrahedron
(FDT).  However  the difference in binding energy was rather small
(of  the order of 0.1 eV).  So we decided to study  the dependence
of the structure of these clusters  on the ionic charge.   Another
reason to look whether the geometry  depends  on the  ionic charge
is that we   found  in I from   our   semi--empirical      quantum
calculation  on these clusters that the charges  on  the ions  are
somewhat smaller than the full ones.

Furthermore,       we   note  that  there  are two structures  for
crystalline LiGe~\cite{LiSi-LiGe}: one equivalent  to  LiSi  (a Si
network  with a three--fold coordination)   and a  second one with
weakly coupled    layers  of  Ge,    with half of the   Ge with  a
four--fold coordination  and the  other  Ge   with   a   two--fold
coordination.    This would lead to charge   imbalance,  with  one
neutral Ge and one Ge$^{2-}$.    An analysis shows that the latter
Ge has two   rather  short interlayer Ge--Ge bonds, which have   a
bondstrength~\cite{bondorder}  of 0.25 each, so its formal valence
charge   is  Ge$^{1.5-}$,    thus   satisfying    the  generalized
Zintl-Bussmann-Klemm valence rule~\cite{Zintl}.   The other alkali
monotetralides   crystallize   either  in  a structure   with well
defined M$_4$  tetrahedra (all Na, K, Rb, Cs monotetralides) or in
a structure with M layers (LiSn) or in a distorted CsCl  structure
(LiPb).  These structures are  reviewed  in~\cite{review     Zintl
solid 1,review Zintl solid  2}.   The  only  known IR spectra  are
reported in~\cite{Schnering_et_al_(1986)}.

This is not the only known   structure  where the tetralide  M$^-$
ions form layered,  three--dimensional network or a structure with
isolated  tetrahedrons.  Earth-alkaline   ditetralides crystallize
all in a structure     with  three--fold   coordinated   tetralide
sublattice~\cite{alkaline-earthditeteralides}       in  a  layered
structure (CaSi$_2$, High-Pressure-High-Temperature   BaSi$_2$) or
in a structure   with tetrahedra (BaSi$_2$,  CaGe$_2$,   SrGe$_2$,
BaGe$_2$),  or in a structure   with a three--dimensional  network
(SrSi$_2$).   Large   cations  seem   to favor the structure  with
tetrahedra, small cations the layered structure.

We choose to study the structure of Li$_{4}$Ge$_{4}$
as a function    of polarizability and ionic charge.  
In the  next  section we give a brief  account  of the theoretical
model.  In section  3 we give  results,  in  section  4 we give  a
discussion of these results and the conclusions.

\section{THEORY.}\label{section: theory}
The theoretical   model  we  use is the same as in I. It is   a mixture  of a
semi--empirical quantum  mechanical  model  and  a classical    electrostatic
model. The  semi--empirical model is used to describe the valence  electrons.
It is basically  an adaptation  for clusters  of the parameterization  scheme
given by Harrison~\cite{SST,Harrison    new}  for  the  hybridization  matrix
elements he derived for  the solid  state.   We have  applied    his  new
scheme~\cite{Harrison new}  and  parameter   values,  which includes  overlap
matrix  elements  calculated    by the  Wolffberg--Helmholtz   approximation.
Electron--electron  interactions     are   neglected   in this   scheme.  The
unperturbed        atomic  orbital   energies    are    also   taken     from
Harrison~\cite{Harrison   new}. We include the peripheral  s state correction
as a perturbation. We include    only the atomic s and p levels of the alkali
and tetralide.

The classical electrostatic contributions to the binding  energy  consist  of
the Coulomb interactions of  the ionic charges ($Q$), the interaction between
the  Coulomb fields    and   the induced polarizability,   the  dipole-dipole
interactions, and the  energy required  to create these  induced dipoles. The
polarizability is taken from Fraga~\cite{Fraga}. All results  are reported as
a function of the fraction of these polarization  values.  The polarizability
of A and M are determined by the same fraction of the Fraga value.

Next to this we have two semiclassical contributions: the van  der
Waals interaction and the Born  repulsion.   For the first we take
the form: $C/R^6$.   The  constant   $C$  is approximated   by the
polarizability.   As this contribution  is in general small we did
not optimize  the parameter   $C$.  For the Born repulsion between
two  atoms/ions     1  and  2 a distance    $R$   apart  we  take:
$F(\rho_1+\rho_2)\exp((R_1^0    +  R_2^0   - R)/(\rho_1+\rho_2))$,
where $R^0_i$ is the Born radius of atom $i$, and $\rho_i$  is the
decay length of the  Born repulsion for $R >  R_1^0 + R_2^0$.  The
Born radii  have been fixed so as to reproduce  the neutral atom interatomic
distances (see \cite{fit:-geertsma-2002b} and I) by this model.

We do not take into account  the  contribution due to the covalent
distortion of the valence  electron  density in the calculation of
the polarization contribution. We also neglect the distortion   of
the core charge density due to core--core   overlap    in case  of
short interatomic distances.

We have calculated  the minimum energy using the  Simplex  method,
using all  atom coordinates in the minimization.   In  some cases,
to be  discussed below,   we have   used   structures with a fixed
geometry: NDT and FDT with  two parameters, PAR and  BUT with four
parameters.

\section{RESULTS AND DISCUSSION.} \label{section: results}

In figure~\ref{figure: phase diagram}    we  present the   cluster
structure  phase     diagram for Li$_{4}$Ge$_{4}$.    We find five
regions:   one where  the tetrahedron   of Ge is stable (NDT), one
where Ge$_2$  piars are   stable  (PAR), another where a butterfly
geometry   - a tetrahedron    with two long  opposite bonds --  is
stable (BUT) and a fourth where   isolated ions   of Ge are stable
(FDT).   These structures  are illustrated in figure  \ref{figure:
LiGe structures}. The fifth region for $\alpha > 0.75  \alpha_F$ is
where the A$_4$M$_4$ can always find a configuration  in which the
polarizability  becomes the most   important  contribution  to the
total   energy.    We   illustrate  these     geometries        in
figure:~\ref{figure:   LiGe structures}  for  the Li$_{4}$Ge$_{4}$
cluster.

Let us first   discuss  the   extent    of the  latter  region  of
instability. It has been known for a long time that including  the
polarizability    of   ions in  the  calculation    of the cluster
structure with minimum energy can give rise to divergences in  the
total energy.  Some years ago, Thole~\cite{polarizability  damping
original} discussed the problem of  the  polarization  catastrophe
in Molecular Dynamics  simulations of molecules    due   to  close
appraoch of two molecules. He derived a criterium  for the nearest
approach  of two molecules   in order to avoid such problems:  the
effective polarizability of two polarizable molecules diverges  if
two  molecules appraoch     each   other   closer    than    $R  =
(2\alpha)^{1/3}$.   This  criterium  is independent   of the local
electric fields.  However, using this criterium our systems  would
be stable.

We performed  a calculation for the butterfly/pair geometry of the
total energy,  and derived  a relation between $R$ distance in the
pairs  (or the diagonal,    in   the butterfly)  and  the distance
between the pairs (height of the  butterfly): $H$.  One then finds
the  following  condition   for   divergence   of     polarization
contribution to the total energy:
\begin{equation}
A=\frac{R^{3}}{\alpha}=\frac{1}{4}\left[x-2+\left(
(3x-2)^{2}+16\frac{H^{2}}{R^{2}}x^{2}\right)^{1/2}\right]
\end{equation}
where  $x=(1/2+(H/R)^{2})^{-5/2}$.     We  find that  for  $R  \le
(4.66\alpha)^{1/3}$,   there  is always an $H$ for which the total
energy diverges.     Using   the values     of Fraga    for    the
polarizability,   we  find   that  the critical    value   of  the
polarization is  $\approx 0.7\alpha_F$.   This is what we actually
find in our simulations.     This limit for  stable  solutions  is
independent of the electric  field.  Obvious within this region of
polarization induced instability there are relative minima.

In this instability  region we find  that  for certain  atoms  the
length of the dipole moment is larger than  the ion diameter.  The
polarization energy  becomes even larger  than the Born repulsion,
and  the ions  can  come very close  together.  Our model  is  not
longer valid: The cores start to overlap  strongly, and one should
also  take into  account   the  deformation    of the core  charge
distribution,  described by  the so--called    deformation  dipole
moment (Tosi et al~\cite{deformation dipole}).

In the region where we find the stable geometries  we can describe
the succession  of structures with increasing polarizability as if
polarizability   tends  to break the  covalent   M-M bonds:  large
polarizability  together with strong electric  fields     act as a
scissor  on  covalent    bonds.   Also from the  point  of view of
increasing  the ionic charges  one finds these  covalent bonds are
broken with increasing ionic charge: from a Ge$_4$ tetrahedron  in
the NDT  structure    with  six Ge--Ge  bonds  with a lenght    of
approximately 2.53  -- 2.61 $\AA$ with the Li far away outside  on
its 4 faces, to a BUT structure with a Ge$_4$ butterfly  with four
short  Ge--Ge  bonds (2.45-2.53 $\AA$),  and two long ``bonds'' of
about  3.00 $\AA$.   This   geometry     is for  relative    large
polarizability  an intermediate  structure to a PAR structure with
two  Ge$_2$ pairs, with  a total of 2 Ge--Ge with bond  lengths of
2.39 -- 2.45   $\AA$.  These pairs are  about  3.68  -- 3.80 $\AA$
apart.  For relative   large  polarizability the Ge  pairs further
dissociate to isolated Ge ions (FDT), where one finds the  four Li
on or just outside the  faces of the large Ge tetrahedron. In this
case the interatomic Ge distance is about 3.7 $\AA$.

In the stable regime we have applied  the Simplex  to minimize the
total energy with all atom coordinates.  In some cases  one has to
be very near to the minimum otherwise the Simplex gets stuck in  a
relative   minimum.  This  happens especially   in the case of two
pairs.  One  of the problems is that there are two  solutions with
pairs in the case  of Li$_{4}$Ge$_{4}$,  which are separated by an
energy of about  0.7 eV.    The  one with highest  energy    has a
somewhat larger  Ge--Ge distance in the Ge$_2$ pairs. Secondly, in
case  of the pairs  with lowest energy, the  energy minimum in the
coordinate   ``landscape''   seems   to be very  localized,    and
difficult     to  find. Such a problem did  not    arise  for   the  other
geometries of Li$_{4}$Ge$_{4}$.

We   have calculated     the  vibration  energies (which we report
elsewhere \cite{fit:-geertsma-2002b}) of these clusters. We find that  the cluster with two
pairs with  the  short Ge--Ge bond has one mode with a very  large
energy  of the order of 0.8 eV. We noted  that  the   Simplex  has
difficulties finding the minimum,  and actually one finds a number
of saddlepoints which   are very close    in coordinate space.  
Note also the large extend of the region in the phase diagram
where the Ge$_2$ pairs are metastable \ref{figure: phase diagram}B.

We find  that    the  ionic   charges    as calculated   from  the
semi--empirical    molecular orbital part of  our  calculation are
approximately 0.8 for all A$_4$M$_4$ clusters.

In the solid  state  we  identify  the M$_4$   tetrahedron binding
energy  with the energy    difference   between  the NDT   and FDT
configuration.    We  calculated  the energy  difference between a
A$_4$M$_4$ cluster with  a M$_4$ tetrahedron  (NDT)  and with four
isolated M ions (FDT).   For the charge  transfer we take  $Q=0.8$
and for the polarizability   $\alpha =\alpha_F$.  The  results are
in table~\ref{table:   energies},   and where they can be compared
with  results  from an analysis of lattice   energies~\cite{review
Zintl solid  1} and  with M$_4$ binding  energies   derived   from
specific heat data~\cite{specific heat}.  Note the good  agreement
of the energy differences  calculated within  our simple model and
the    binding    energies   obtained   from the  specific    heat
data~\cite{specific          heat}      and     Madelung    energy
analysis~\cite{review Zintl solid 1}.

\section{DISCUSSION AND CONCLUSIONS.
\label{section:discussion-and-conclusion}}    

Using    this   relative    simple   model,     introduced      in
section~\ref{section: theory},    for the calculation of the total
energy of ionic-covalent clusters,  we have studied  the influence
of the polarizability on the geometry  and structure of a relative
simple  cluster  Li$_{4}$Ge$_{4}$.   We  find that  polarizability
together    with ionic  charge can break covalent bonds.   This is
contrary  to the     usual   believe that  polarization      plays
approximately    the    same  role  as  covalent bonding.   Strong
polarizability   breaks    covalent bonds.  In order to access the
validity  of our model  for these A$_4$M$_4$     clusters  we also
calculated     the     dissociation        energy   and  vibration
energies~\cite{fit:-geertsma-2002b}   of these clusters.     Where
experimental   data are  available   we find good agreement   with
experiment.

A region  of  instability in the phase diagram of Li$_{4}$Ge$_{4}$
clusters  is found for large  polarizability.  The most simple way
to remedy    the deficiency   of our   model is  to  introduce the
deformation  dipole,   which   develops   when two  atom cores are
overlapping. 

From the  phase diagram of LiGe  \ref{figure:  phase diagram}A   we
conclude     that for small   anion   polarizability         these
Li$_{4}$Ge$_{4}$ clusters have a cubic  structure  with anions and
cations approximately on  the vertices  of a cube, independent  of
the ionic charge.  For increasing anion polarizability  the Ge$_4$
tetrahedron  deforms into  two pairs, separated   by   the four Li
cations, or into a butterfly, with two relative  long M--M  bonds,
and four normal M--M bond lengths,  or into  a structure where all
the covalent   bonds of the  Ge$_4$   tetrahedron  are  broken up.
Further increasing  the  polarizability  we enter  into  a  regime
where no structure is stable. In case  of Sn and Pb we do not find
these intermediate phase of pair or butterfly.

Let us  now briefly  comment on how  we can  apply this results to
solid state  structures.   In general  we expect the local Coulomb
fields  in  a solid  to be smaller than in the clusters   we  have
studied.   In order  to  apply  our results  to the solid state we
have to keep  this in mind. Firstly, for small cations like Li and Mg
we donot expect  isolated M$_4$  tetrahedra in the solid  state --
unless when $\alpha_M$  is very small.   A possible characteristic
for  the structure of these compounds  are in this case: a 3--fold
coordinated 2d or  3d network   of the M sublattice  or a strongly
distorted ionic structure (NaCl, CsCl).

Secondly, For  large cations   like  K, Rb,  Cs, Ba one expects based on the
calculations presented in this  paper  isolated M$_4$ tetrahedra.
Na,  Ca  and    Sr   are  intermediate.     In  our  very    first
paper~\cite{paper      I}  on this subject  we  applied   a simple
hybridization model to a  modified Bethe lattice.  We found a very
clear  separation  between the isolated tetrahedron structures  of
Na,  K, Rb and Cs mono tetralides and the Li monotetralides.   The
different   structures   found for these Li compounds  can only be
explained by taking into account polarizability.

It is clear  that one finds the same trends in  the solid state as
found   for  these small    clusters.  The  pair    and  butterfly
configuration can be compared with the 2--dimensional layered  and
3--dimensional network structures found for LiGe and LiSi.

Next let us  turn  our attention   to the  liquid state  of  these
alkali  monotetralides.         For    reviews    of this    field
see:~\cite{review  liquid}.    LiSn and LiPb are not  of  interest
because the tetrahedra  are  already broken up in the solid state.
In the liquid  state of  Cs, Rb and possibly K monotetralides  one
finds tetrahedra, which,  due to entropic effects, break up with increasing  temperature into
smaller units.   This holds  probably also
for NaSn  and NaPb.   This is actually  the interpretation  of the
Schottky   anomaly   found  in  the   liquid   state     of  these
systems~\cite{specific   heat}. In case of NaSi and NaGe  there is
also  the possibility    of  the  formation of  3-fold coordinated
network: in  our calculations the BUT  and PAR configurations  are
metastable  phases with an energy not  far  above   the  NDT. This
offers also an explanation for  the occurance  of these M networks
observed   in MD~\cite{CarPar,MolDyn}        and  Reverse    Monte
Carlo~\cite{RevMonCar} simulations.

We  conclude  that    these model calculations         on    small
ionic--covalent     clusters can   give us reliable  data  on  the
interatomic   distances, binding energy and vibration  spectra  of
such small   clusters,   and  is  a basis   for  the discussion of
structural   phase   transitions  in  alkali-monotetralides    and
alkaline--earth--ditetralides.

\section*{Acknowledgements}
We acknowledge a grant from the Europea Union ( TMR contract:
ERBFMBICT--950218), which made the first stages of this research possible,
and a grant from CNPq (300928/97--0) during the final stages of this work.

\vfill
\newpage 
\pagebreak
\begin{table}[h]
\caption[table 1]{The binding energy (eV) of tetralide tetrahedra. A:
derived from the energy difference between the NDT and FDT geometry as
described in the main text. B: derived from the best fit to specific heat
data (see \cite{specific heat}). C: Derived from lattice energy
difference taken from \cite{review Zintl solid 1}.\label{table: energies}} 
\begin{center}
\begin{tabular}{c ccc ccc}
\hline
\hline
     & KSn   &  RbSn  &  CsSn   & KPb   &    RbPb   &  CsPb \\
\hline
A    & 1.8   &  2.4   &  2.5    &  0.9  &   1.5     &  1.7  \\
B    & 1.8   &  2.0   &  2.0    &  1.6  &  1.5$^5$  &  1.7  \\
C    & 1.7   &  1.80  &  1.97   &  1.4  &  1.54     &  1.67 \\
\hline
\hline
\end{tabular}
\end{center}
\end{table}
\vfill
\newpage
\begin{figure}[h]
\caption[Fig:01]{In   this cluster   phase  diagram (A)  we  give the
various stability  regimes of  Li$_{4}$Ge$_{4}$  as a function  of
ionic charge  ($Q$)  and polarizability   $\alpha/\alpha_F$.   The
dotted line in the BUT region separates the metastable    PAR from
metastable NDT in the  BUT region  of the   phase diagram.  In the
B we  show the extend of the metastable   PAR phase.  For more
details see the main text.\label{figure: phase diagram}}
\end{figure}
\begin{figure}[h]
\caption[Fig:02]{The Li$_4$Ge$_4$ cluster structures:  NDT; PAR; BUT; IDT.
\label{figure: LiGe structures}} 
\end{figure}
\vfill
\pagebreak
\epsfverbosetrue
\epsfysize=4in
\epsffile{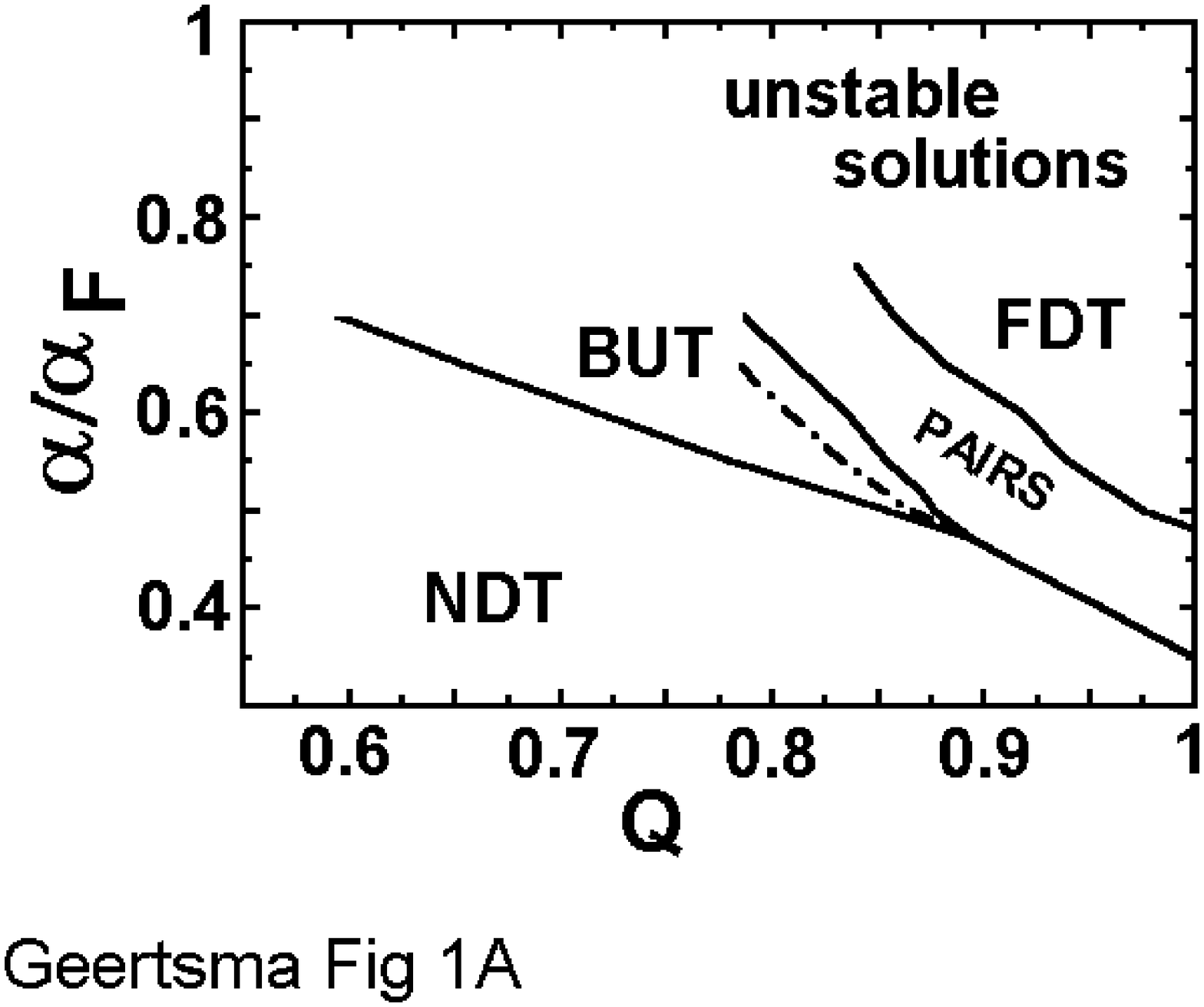}
\epsfysize=4in
\epsffile{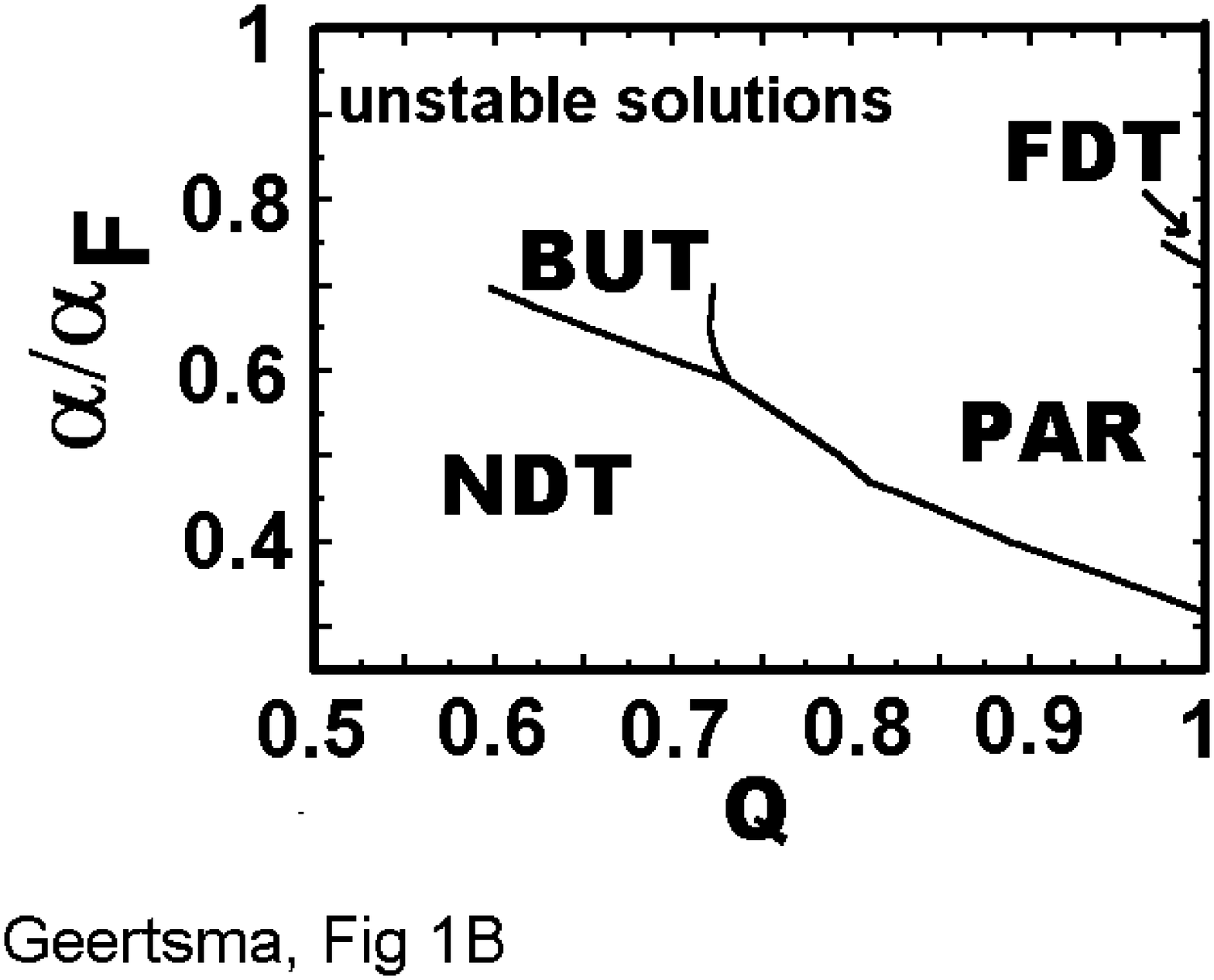}
\epsfysize=4in
\epsffile{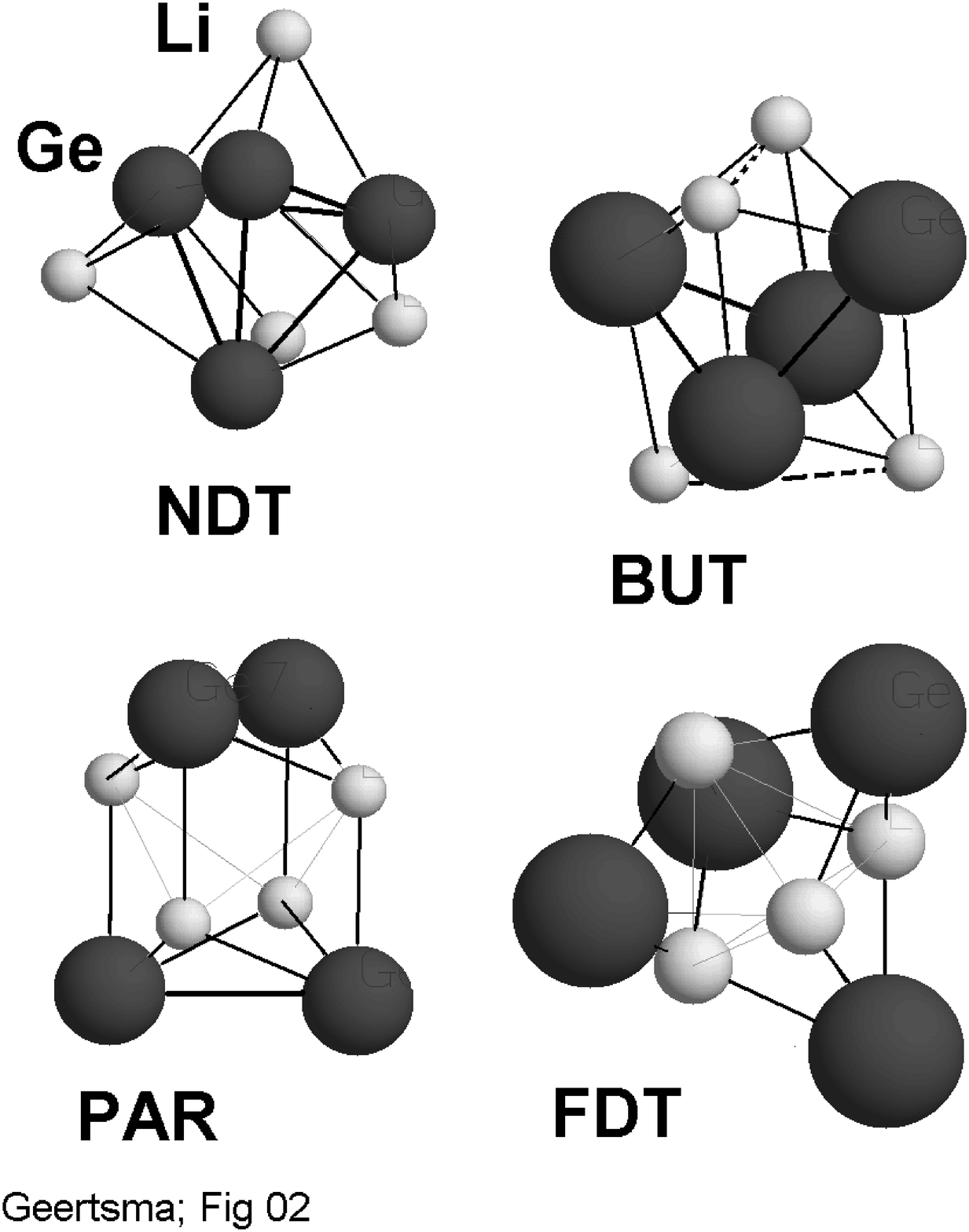}


\newpage
\begin{thebibliography}{10}
\bibitem[1]{geertsma 2002}W.~Geertsma, to be submitted to J.
Phys.: Condens. Matter..

\bibitem[2]{Fraga}  S.~Fraga and J.~Muszynska, {\it Physical Science Data}
Vol. 8, {\it Atoms in External Fields}, (Elsevier Scientific Publ. Comp. (1981)).

\bibitem[3]{LiSi-LiGe}
J.~Evers and G.~Oehlinger and G.~Sextl,
Angew. Chem. Int. Ed. Engl., {\bf  32} (1993) 1442--1443;
J.~Evers and G.~Oehlinger and G.~Sextl and H.--O.~Becker,
Angew. Chem. Int. Ed. Engl., {\bf  26} (1987) 76, and
P.~Sherwood and R.~Hoffmann,
J. Am.  Chem. Soc.,
{\bf  112} (1990) 2881-2886.
\bibitem[4]{bondorder}An often used emperical relation between distance and
bond order is $v = 10^{(R_0-R)/A}$ with $R_0$ the single bond length value
and $A=0.8$, see for example: I. D. Brown, Chem. Soc. Rev. {\bf
7} (1978) 359--376. 
\bibitem[5]{Zintl}  I.~F.~Hewaidy, E.~Bussmann and W.~Klemm, Z. anorg.\ allg.\
Chem., {\bf 327} (1964) 283--293.
\bibitem[6]{review Zintl solid 1}  H.~G.~von Schnering, Angew. Chem. Int.
Ed. Engl. {\bf 20} (1981) 33--51.
\bibitem[7]{review Zintl solid 2}  R.~Nesper, Progr. Solid State Chem. {\bf %
20} (1990) 1--45.
\bibitem[8]{Schnering_et_al_(1986)}
H.~G.~von Schnering and  M.~Schwartz and R.~Nesper, Angew. Chemie,
{\bf 98 } (1986) 558--559.
\bibitem[9]{alkaline-earthditeteralides}
K.~H.~Janzon and H.~Sch\"afer and A.~Weiss, Z. Naturforsch b,{\bf 23      }
(1968) 1544; 
J.~Evers and G.~Oehlinger and A.~Weiss, J. Sol. State Chem.,{\bf
20}, (1977) 173--181;
J.~Evers and G.~Oehlinger and A.~Weiss, Z. Naturforsch. b, {\bf  37     }
(1982) 1487--1488;
J.~Evers and G.~Oehlinger and A.~Weiss, Z. Naturforsch. b,
{\bf33       } (1978) 956;
H.~J.~Wallbaum, Naturwissenschaften,
{\bf    32   } (1944) 76;
A.~Betz and H.~Sch\"afer and A.~Weiss and R.~Wulf, Z. Naturforsch. b,
{\bf  23}, (1968) 878.
J.~Evers and G.~Oehlinger and A.~Weiss, Z. Naturforsch. b,
{\bf 34      } (1979) 524.
J.~Evers and G.~Oehlinger and A.~Weiss, Z. Naturforsch. b,
{\bf  32     } (1977) 1352--1353.
J.~Evers and G.~Oehlinger and A.~Weiss, Z. Naturforsch.  b,
{\bf   35    } (1980) 397--398.
\bibitem[10]{SST}  W.~A.~Harrison, {\it Electronic structure and the properties
of solids}, (San Fransisco: W.H. Freeman (1980)).
\bibitem[11]{Harrison new}  W.~A.~Harrison, Phys. Rev. B {\bf 24} (1981)
5835--5843, and Phys. Rev. B, {\bf 27} (1983) 3592--3604.
\bibitem[12]{fit:-geertsma-2002b}W.~Geertsma, In preparation.
\bibitem[13]{polarizability damping original}B.~T.~Thole, Chem. Phys., {\bf 59
} (1981) 341, and   P.~Th.~van Duijnen and M.~Swart,  J.  Phys. Chem.  A,{\bf
102} (1998) 2399--2407.
\bibitem[14]{deformation dipole}Z.~Akhdeniz and M.~P.~Tosi, Phys. Chem Liq., 
{\bf 17} (1987)  91--104.
\bibitem[15]{paper I}  W.~Geertsma, H.~Dijkstra and W.~van der Lugt, J. Phys. F 
{\bf 14} (1984) 1833--1845.
\bibitem[16]{specific heat}  W.~Geertsma and M.--L. Saboungi, J. Phys. Cond.
Matter, {\bf 7} (1995) 4803--4820.
\bibitem[17]{review liquid}  M.--L.~Saboungi, W.~Geertsma and
D.~L.~Price, Annu.
Rev. Phys. Chem., {\bf 41} (1990)  207--244, and 
W.~van der Lugt, J. Condens.
Matter {\bf 8} (1996) 6115--6138.
\bibitem[18]{CarPar}  G.~A.~de Wijs, G.~Pastore, A.~Selloni and 
W.~van~der~Lugt,  Phys. Rev. B {\bf 48} (1993) 13459--13468;  
G.~Galli and M.~Parrinello,
J. Chem. Phys.,  {\bf  95} (1991) 7504--7511;  
G.~A.~de ~Wijs, G.~Pastore, 
A.~Selloni and W.~van der Lugt,  Europhys. Lett. {\bf 27} (1994)  667--672,  and
ibid. J. Chem. Phys. {\bf 103}  (1995)   5031--5040; 
G.~Seifert and G.~Pastore  and R.~Car, J. Phys.: Condens Matter, {\bf 4 } (1992) L179--L183.
\bibitem[19]{MolDyn}   M.~Sch\"{o}ne   and   R.~Kaschner       and
G.~Seifert, J.  Phys.   Condens.  Matter, {\bf 7}  (1995) L19-L26;
J.~Hafner,  K.~Seifert--Lorenz  and O.~Genser, J. Non--Crystalline
Solids, {\bf 250--252}  (1999)  225--235,  {\it and} O.~Genser and
J.~Hafner, J. Non--Cryst.  Sol.,  {\bf  250--252} (1999) 236--240;
A.~Senda, F.~Shimoji and K.~Hoshino, J. Non--Cryst.    Sol.,  {\bf
250--252} (1999) 258--262.
\bibitem[20]{RevMonCar}   M.~Stolz,    R.~Winter,   W.~S.~Howells,
R.~L.~McGreevy,     J. Phys.:  Condens.     Matter {\bf  7} (1995)
5733-5743., {\it and} M.~Stolz, O.~Leichtweiss, R.~Winter,  M.--L.
Saboungi, J.~Fortner and W.~S.~Howells, Europhys.   Lett. {\bf 27}
(1994)  221--226, {\it and} P.~H.~K.~de Jong, P.~Verkerk, L.~A.~de
Graaff, W.~S.~Howells  and W.~van~der~Lugt, J.    Phys.:  Condens.
Matter   {\bf 7} (1995)  499--516,  {\it  and} P.~H.~K.~de   Jong,
P.~Verkerk,W.~van der  Lugt  and L.~A.~de  Graaff, J.  Non--Cryst.
Solids {\bf 156--158} (1993) 978--981.
\end{thebibliography}
\end{document}